\documentclass[amsmat,amssymb,amsfonts,aps,prb,twocolumn,showpacs]{revtex4}
\usepackage{graphicx}
\usepackage{dcolumn}
\usepackage{bm}
\newcommand{\beq}{\begin{equation}}  
\newcommand{\eeq}{\end{equation}}  
\newcommand{\beqa}{\begin{eqnarray}}  
\newcommand{\eeqa}{\end{eqnarray}}  
  
\newcommand{\ome}{\omega}

\begin{document}

\title{Cotunneling theory of   inelastic STM  spin  spectroscopy}

\author{F. Delgado, and J. Fern\'andez-Rossier}
\affiliation{Departamento de F\'{\i}sica Aplicada,
Universidad de Alicante, San Vicente del Raspeig, 03690 Spain}

\date{\today}

\begin{abstract}
We propose cotunneling as the microscopic mechanism that makes possible inelastic electron spectroscopy of magnetic atoms in surfaces for a wide range of systems, including single magnetic adatoms, molecules and molecular stacks.  We describe electronic transport between  the scanning tip and the conducting surface through the magnetic system (MS) with a generalized Anderson model, without making use of effective spin models. Transport and
spin dynamics are described with an effective cotunneling Hamiltonian in which the  correlations in the magnetic system are calculated exactly
and the coupling to the electrodes is included up   to second  order in the tip-MS and MS-substrate.
In the adequate limit our approach is equivalent to the phenomenological  Kondo exchange model that successfully describe the experiments .
  We apply our method to study in detail inelastic transport in  two systems, stacks of Cobalt Phthalocyanines and a single Mn atom on Cu$_2$N.  
  Our method accounts both, for the  large contribution of the inelastic spin exchange events to the  conductance and
   the  observed conductance asymmetry.
\end{abstract}

 \maketitle

\section{Introduction\label{intro}}
The combination of two powerful techniques, 
Inelastic Electron Tunneling spectroscopy (IETS) and Scanning Tunneling Microscope (STM) makes it possible to probe inelastic excitations with subatomic resolution. The STM-IETS technique was first applied to the study of vibrational excitations  of single molecules on surfaces \cite{Stipe_Rezaei_science_1998} and has more recently been used to  study  spin excitations of  a single and a few magnetic atoms and molecules deposited on surfaces.\cite{Heinrich_Gupta_science_2004,Hirjibehedin_Lutz_Science_2006,Hirjibehedin_Lin_Science_2007,Otte_Ternes_natphys_2008,Loth_Bergmann_natphys_2010,Loth_Lutz_njphys_2010,Chen_Fu_prl_2008,Tsukahara_Noto_prl_2009,Fu_Zhang_prl_2009,Fu_Ji_apl_2009,Khajetoorians_Chilian_nature_2010,Loth_Etzkorn_Science_2010} 
In STM-IETS,   electrons tunnel between the tip and the conducting substrate going through the magnetic system. As the bias voltage $V$ is increased, a new conduction channel opens whenever $eV$ is larger than the energy of some internal excitation of the atom, which results in a stepwise  increase of the differential conductance $dI/dV$ and a peak or dip in the $d^2I/dV^2$.  Tracing the evolution of the elementary excitations as a function of an applied magnetic field and fitting to effective spin Hamiltonians  permits  to infer the single ion magnetic anisotropy tensor as well as exchange coupling between adjacent atoms and molecules.\cite{Heinrich_Gupta_science_2004,Hirjibehedin_Lutz_Science_2006,Hirjibehedin_Lin_Science_2007,Otte_Ternes_natphys_2008,Loth_Bergmann_natphys_2010,Loth_Lutz_njphys_2010,Chen_Fu_prl_2008,Tsukahara_Noto_prl_2009,Fu_Zhang_prl_2009,Fu_Ji_apl_2009,Khajetoorians_Chilian_nature_2010}
 
 The IETS-STM technique has been applied to a variety of magnetic systems weakly coupled to a conducting substrate. The list includes a single transition metal  atom (Mn, Fe,  Co) deposited on a single monolayer of Cu$_2$N on Copper \cite{Hirjibehedin_Lin_Science_2007,Otte_Ternes_natphys_2008,Loth_Bergmann_natphys_2010,Loth_Lutz_njphys_2010,Loth_Etzkorn_Science_2010}, to chains of up to 10 Mn atoms on the same substrate\cite{Hirjibehedin_Lutz_Science_2006}, to Fe-phthalocyanine  (Fe-PC) molecules on oxidized Cu \cite{Tsukahara_Noto_prl_2009},  to stacks of Co-PC molecules on Pb\cite{ Chen_Fu_prl_2008,Fu_Zhang_prl_2009}, to Mn-PC on PbO \cite{Fu_Ji_apl_2009} and, more recently,  a single Fe atom on InSb, a semiconducting substrate.\cite{Khajetoorians_Chilian_nature_2010} 
 For all these systems it is possible to describe the spin exchange  assisted tunneling, which  accounts for the coupling  between transport electrons and the localized spins of the magnetic atoms or molecules, with Kondo-like Hamiltonians.\cite{Appelbaum_pr_1967,Rossier_prl_2009,Fransson_nanolett_2009,Persson_prl_2009,Delgado_Palacios_prl_2010,Fransson_Eriksson_prb_2010,Zitko_Pruschke_njphys_2010,Sothmann_Konig_njp_2010,Delgado_Rossier_prb_2010}  Whereas this approach  successfully describes the main  experimental results, including the differential conductance, as well as  effects related to   current driven spin dynamics and/or spin polarized tip, there are questions that can not be addressed using effective spin models: 
\begin{enumerate}
\item Why the spin assisted inelastic conductance is comparable to the elastic contribution, in contrast with the phonon-assisted inelastic contribution?
\item What is the microscopic origin of the spin exchange tunneling?
\item Why the inelastic conductance is not always symmetric with respect to the inversion of the  bias polarity?
\end{enumerate}

In this work we provide a theoretical framework to model the existing STM-IETS experiments that  addresses these questions. Our starting point is a generalized multi-orbital/multi-site Anderson model, in which the electrons in the localized orbitals of the magnetic system (MS) are hybridized to the itinerant states of the tip and the surface. The states of the MS are calculated by exact diagonalization of a microscopic Hamiltonian that can include Coulomb repulsion, crystal field and spin-orbit coupling. 
 Transport and spin dynamics are described by means of an {\em effective cotunneling Hamiltonian} in which the coupling to the tip and surface is included up to second order.  This approach works provided that  the charging energy of the MS is much larger than the temperature, applied bias potential and the electrode induced broadening of the MS levels.   Thus, the MS must be in  the Coulomb Blockade situation,  where the charge is a good quantum number and
 current flows due to  quantum  charge fluctuations, known as cotunneling.\cite{Averin_Nazarov_prl_1990,Averin_Nazarov_book} 

When applied to a single orbital Anderson model,  the effective cotunneling  Hamiltonian that we obtain is identical to the Kondo model obtained through the standard Schrieffer-Wolff transformation.\cite{Anderson_prl_1966,Schrieffer_Wolff_pr_1966}  Our method can applied to systems with more than one localized orbital,   necessary to address most experimentally relevant systems.
\cite{Heinrich_Gupta_science_2004,Hirjibehedin_Lutz_Science_2006,Hirjibehedin_Lin_Science_2007,Otte_Ternes_natphys_2008,Loth_Bergmann_natphys_2010,Loth_Lutz_njphys_2010,Chen_Fu_prl_2008,Tsukahara_Noto_prl_2009,Fu_Zhang_prl_2009,Fu_Ji_apl_2009,Khajetoorians_Chilian_nature_2010,Loth_Etzkorn_Science_2010}
The effective cotunneling Hamiltonian describes transitions between the different many-body states of the MS induced by their coupling to the itinerant electrons. 
  This permits to calculate the scattering rates,  both for the dissipative dynamics of the spin excitations of the MS coupled to the leads and those leading to the current.  
  
The rest of the paper is organized as follows. In Sec. II we present the derivation of the effective Hamiltonian and the procedure used to calculate the current, leaving some of the technical details for the appendix. In Sec. III we apply our approach to the case of a single site Anderson model, which permits to test our approach against well established results. In Sec. IV  we implement our approach to model transport through stacks of CoPc molecules\cite{Chen_Fu_prl_2008,Fu_Zhang_prl_2009}. For that matter, we describe the CoPc stacks by means of a Hubbard model.  In Sec. V we study the case of a single Mn adatom on a Cu$_2$N surface,\cite{Loth_Bergmann_natphys_2010,Hirjibehedin_Lutz_Science_2006} 
using a multi-orbital Anderson model where 
Coulomb interaction, crystal field and spin-orbit coupling in the MS  are  included in the  Hamiltonian and treated exactly, by means of numerical diagonalization. In section VI we summarize our main results.

%
\section{Theory}
\subsection{Effective Hamiltonian\label{theory}}
%
%
%
We describe a magnetic system weakly coupled to two electrodes, denoted as   tip ($T$) and surface ($S$) without loss of generality,   using the following
Hamiltonian:
 \begin{equation}
{\cal H}= {\cal H}_{\rm T} + {\cal H}_{\rm S} + {\cal H}_{\rm MS}
+{\cal V}_{\rm tun}.
\end{equation}
Here ${\cal H}_{\rm T} + {\cal H}_{\rm S}$ correspond to the Hamiltonian of the two electrodes, ${\cal H}_{\rm MS}$  the magnetic system and ${\cal V}_{\rm tun}$ the tunneling Hamiltonian. We shall consider the two electrodes as free electron reservoirs, i.e., 
${\cal H}_{\rm T} + {\cal H}_{\rm S} =  \sum_{\alpha} \epsilon_{\alpha} f^{\dagger}_{\alpha}  f_{\alpha}$, 
where $f_{\alpha}^\dag$ ($f_{\alpha}$) is the creation (annihilation) operator of a quasiparticle with single particle number $\alpha\equiv \left\{k,\eta,\sigma\right\}$, with
 momentum $k$,  electrode $\eta=T,\;S$ and spin projection  in the quantization direction $\sigma$.
In general, the central region has a complicated many-body  Hamiltonian that includes Coulomb repulsion, spin-orbit coupling, crystal field terms and so on.   The many-body eigenstates of ${\cal H}_{\rm MS}$, $|q,n\rangle$   have a well defined number of electrons $q$. Only   3 charge sectors, $q=q_0$, $+$ and $-$, are relevant. The $q_0$  corresponds to the ground state of the MS. The sectors $+$ and $-$ correspond to the MS with an extra electron ($q_0+1$) and an extra hole ($q_0-1$) respectively.
The Hamiltonian of the isolated MS can be written as:
\begin{equation}
{\cal H}_{\rm MS} =  \sum_{q,n}  E_{q,n} |q,n\rangle\langle q,n|. 
\label{hms}
\end{equation}
  The tunneling Hamiltonian is given by 
\beqa
{\cal V}_{\rm tun} =\sum_{{\bf i},\alpha} V_{\alpha,{\bf i}} 
 f^{\dagger}_{\alpha} d_{{\bf i}}+h.c.
=\hat{\cal V}^{-}+
 \hat{\cal V}^{+},
 \label{tunnel}
\eeqa
where  the tunneling of electrons in and out the MS are described by   $\hat{\cal V}^{+}$ and
$\hat{\cal V}^{-}$ respectively.
Here $d_{{\bf i}}^\dag$ ($d_{{\bf i}}$) are the creation (annihilation) operator of an electron in a single particle state ${\bf i}\equiv \{i,\sigma\}$ with orbital quantum numbers $i$ and  spin $\sigma$. We assume that single-particle tunneling events are spin conserving and spin independent, i.e.,  $V_{k\eta\sigma,i\sigma'}=V_{k\eta,i}\delta_{\sigma,\sigma'}$.

\begin{figure}[t]
\begin{center}
\includegraphics[angle=0,width=0.9\linewidth]{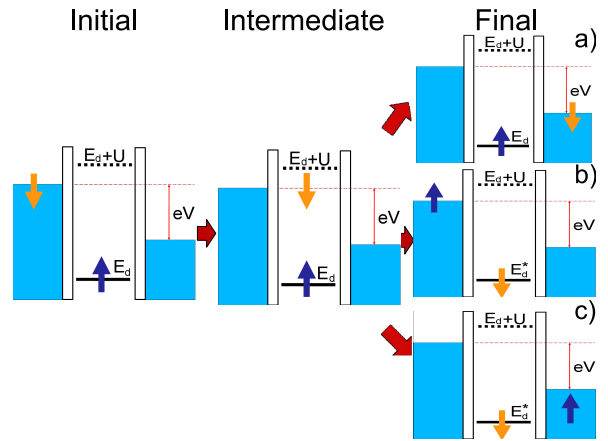}
\end{center}
\caption{ \label{fig2}(Color online) Scheme of the cotunneling transport process through an almost degenerate multiplet of orbital levels far from resonant. 
(a)Elastic transport process without change in the magnetic system  state. (b) Inelastic excitation/relaxation process leading to the creation of an electron-hole pair in one of the electrodes and no charge transport. (c) Inelastic transport process with change of the magnetic system  state. In cases (b) and (c), the energy of the new configuration of the magnetic system changes from $E_d\to E_d^*$. 
}
\end{figure}

We will start with the  uncoupled Hamiltonian 
${\cal H}_0= {\cal H}_{\rm T} + {\cal H}_{\rm S} + {\cal H}_{\rm MS}.$
Since ${\cal H}_0$ commutes with the charge operator of the MS,  the eigenstates of ${\cal H}_0$  can be labeled according to the charge $q$ in the central atom.  We assume that the eigenvalues  in the $q_0$ sector are separated by a large gap from the states in the $q=\pm$ sectors, see Fig.~\ref{fig2}.  In particular, 
 the  chemical potentials of the MS, defined as 
 $\mu_h=E_{G}(N_e)-E_{G}(N_e-1)$ and $\mu_e=E_{G}(N_e+1)-E_{G}(N_e)$, with $E_G(N_e)$ the ground state energy corresponding to $N_e$ electrons, must satisfy 
 $|\mu_h-\mu_\eta |,\;|\mu_e -\mu_\eta|\gg k_bT,|eV|$.
 This corresponds to the conditions of {\em deep cotunneling} in which the sequential-first order transitions are exponentially suppressed.\cite{Recher_Sukhorukov_prl_2000,Wegewijs_Nazarov_arXiv_2001,Bas_Aligia_jphys_2010}
In this limit we can use degenerate perturbation theory to determine the dynamics of the states in the $q_0$ sector, which we denote  with  $|N\rangle$. These states are tensor products of the electrode ground states and the many body states $|q,n\rangle$ of the magnetic system.  The tunneling  operator (\ref{tunnel}) connects them to 
states $|M_\pm\rangle$ that are products of electrode states with 1 quasiparticle and MS states $|q_0\pm 1, m\rangle$. 
Unless otherwise stated, in the rest of the paper we label the MS islands  states with the shorthand notation $|n\rangle\equiv|q_0,n\rangle$ and $|m_{\pm}\rangle=|q_0\pm 1, m\rangle$.

Using degenerate perturbation\cite{Messiah_book_1999} theory we can obtain an effective Hamiltonian for the $q_0$ sector where the tunneling events are included to the lowest order: 
\begin{eqnarray}
 {\cal H}_{\rm cotun}
&=&\sum_{M_\mp} 
\frac{{\cal V}^\pm |M_\mp\rangle\langle M_\mp |{\cal V}^\mp }{E_{M_\mp}-E_0}.
\label{pert}
\end{eqnarray}
In the calculation of the effective Hamiltonian  we are neglecting the  energy variations of the unperturbed states inside the $q_0$ manifold, all taken to be $E_0$,  compared to the charging energy.  
When expanding this operator in the basis of  the electrode quasiparticles and the MS  many-body states, we can
write the effective Hamiltonian for the $q_0$ sector as (see Appendix \ref{appendixA} for details):
\beqa
 {\cal H}_{\rm cotun}=\sum_{\alpha\alpha'}\left[
\hat{{\cal O}}^{(+)}_{\alpha\alpha'}
 -\hat{{\cal O}}^{(-)}_{\alpha'\alpha}\right]  f^{\dagger}_{\alpha}f_{\alpha'}+\sum_\alpha \hat{T}^{(-)}_{\alpha\alpha},
\label{hteffect0}
\eeqa
where   
\begin{eqnarray}
\hat{\cal O}_{\alpha\alpha'}^\pm\equiv \sum_{n,n'}
\langle n|\hat{{\cal O}}^{(\pm)}_{\alpha\alpha'}|n'\rangle |n\rangle\langle n' |
\end{eqnarray}
are operators that act exclusively on the subspace $|q_0,n\rangle$ of the neutral MS. Their matrix elements read:
\begin{eqnarray}
\langle n|\hat{{\cal O}}^{(+)}_{\alpha\alpha'}|n'\rangle=
\sum_{i i',\sigma,\sigma',m_+}
\frac{V_{\alpha, i}V^*_{\alpha', i'}}{E_{m_+}-E_0-\epsilon_{\alpha}}
\gamma_{ n, n'}^{m_+}( ii',\sigma\sigma')
\label{oplus}
\end{eqnarray}
and
\begin{eqnarray}
\langle n|\hat{{\cal O}}^{(-)}_{\alpha\alpha'}|n'\rangle=
\sum_{i i',\sigma,\sigma',m_-}\frac{V_{\alpha, i}^*V_{\alpha', i'}}{E_{m_-}-E_0+\epsilon_{\alpha'}}
\gamma_{n,n'}^{m_-}( ii',\sigma\sigma'),
\label{ominus}
\end{eqnarray}
where
\beqa
\label{gplus}
\gamma_{ n, n'}^{m_+}( ii',\sigma\sigma')&=&
\langle n|d_{i\sigma}|m_+
\rangle\langle m_+|d_{ i'\sigma'}^\dag|n'\rangle
\\
\gamma_{ n, n'}^{m_-}( ii',\sigma\sigma')&=&
\langle n|d_{ i\sigma'}^\dag|m_-
\rangle\langle m_-|d_{ i'\sigma}|n'\rangle.
\label{gminus}
\eeqa

Eqs. (\ref{hteffect0}-\ref{gminus}) constitute the cornerstone of the  formalism. The Hamiltonian ${\cal H}_{\rm cotun}$ in Eq. (\ref{hteffect0}) describes the scattering of a  quasiparticle from the single particle state $\alpha'$ to $\alpha$ in the electrodes together with 
a transition between two many-body states of the MS  within the $q_0$ manifold.  
Three types of elementary processes are described by the effective cotunneling Hamiltonian:   
 elastic processes in which transport electrons are transferred between both electrodes without changes in the central region, creation of electron-hole pair in a given electrode with the corresponding transition in the central island, and inelastic tunneling events.
In all of them, it is apparent from Eqs. (\ref{gplus}) and (\ref{gminus}) that  the excitations within the $q_0$ manifold in the MS occurs via  {\it virtual transitions} to the charged manifolds $q= -$ and  $q=+$.  
An scheme of each of these processes can be seen in Fig.~\ref{fig2}.

Very much like in the case of effective Kondo models, the quasiparticle scattering  events  can be classified in four groups depending on whether they include, or not,  spin flip and/or  electrode transition.   In turn, the spin conserving events are split in two more groups, depending on weather or not they have spin dependent amplitudes. 
Because of the spin rotational invariance imposed in the tunneling Hamiltonian (\ref{tunnel}),  quasiparticle spin flip events imply spin transfer to the MS.  
 Finally, the last term in Eq. (\ref{hteffect0}) describes a renormalization of the many-body levels of the MS and  can be re-adsorbed into a new Hamiltonian for the central part, ${\cal H}_{\rm MS}'={\cal H}_{\rm MS}+\sum_\alpha \hat{T}^{(-)}_{\alpha\alpha}$, so it will be omitted in the following analysis.

 For a fixed set of initial and final quasiparticle states, $\alpha,\alpha'$, the matrices (\ref{ominus}-\ref{oplus})
 have, at most, the dimension of the $q_0$ manifold. 
For instance, as discussed in detail in Sec. \ref{Anderson}  in  so called Anderson model,  when  the states with $q=q_0$ in the island are  those of an unpaired electron,  the dimension of the matrices (\ref{oplus}-\ref{ominus}) is 2, 
corresponding to the two spin projections of a spin $1/2$. As a result, the
Hamiltonian \ref{hteffect0} describes a Kondo coupling between the electrode and the spin $1/2$ of the MS.

\subsection{ Master equation, transition rates  and current\label{rates}}

The procedure described above yields an effective Hamiltonian of the MS coupled to 
the electrodes for which the states of the $q=\pm$ sectors have been integrated out. The effective total Hamiltonian of the electrodes coupled to the $q_0$ manifold reads:
\begin{equation}
{\cal H}_{\rm eff}= \sum_{n}  E_{n} |n\rangle\langle n| + {\cal H}_{\rm T} + {\cal H}_{\rm S} + {\cal H}_{\rm cotun}.
\end{equation}
This Hamiltonian   serves as starting point to calculate both current and 
dynamics of the many body states of the MS within the $q_0$ manifold. The dissipative dynamics of the $n$ states in the MS is induced by the coupling to the electrodes as described by ${\cal H}_{\rm cotun}$.  The master equation for the populations of the MS states, $P_n$,  is given by
\begin{equation}
\frac{dP_n}{dt}= \sum_{n'} W_{n',n} P_n' -P_n\sum_{n'}W_{n,n'},
\label{master}
\end{equation} 
where the transition rates  $W_{nn'}$ for  the MS to go from state $n$ to $n'$  due to quasiparticle scattering in the electrodes are calculated by applying the Fermi Golden Rule with the perturbation given by the tunneling Hamiltonian (\ref{hteffect0}).
 The steady state solutions of this master equation  depend, in general, on the Hamiltonian parameters, the temperature and the bias voltage.  At zero bias, the steady state solutions are those of thermal equilibrium. At finite bias, $P_n(V)$ can depart significantly from equilibrium  depending on the relative efficiency of the transport assisted excitations and relaxations.\cite{Delgado_Rossier_prb_2010}

 The rates $W_{n,n'}$ are the sum of scattering processes in which the initial and final electrode and spin quantum numbers of the  quasiparticle are well defined, 
\begin{equation}
W_{n,n'}
=\sum_{\sigma\sigma',\eta\eta'} W_{n,n'}^{\eta\sigma,\eta'\sigma'}.
\end{equation}
An explicit expression for the spin and electrode dependent scattering rate $W_{n,n'}^{\eta\sigma,\eta'\sigma'}$ is given in the Appendix \ref{appendixA}, Eq. (\ref{gamma}).  The expression involves a convolution over the energy dependent density of states and effective cotunneling rates.  A simpler expression is obtained by doing a number of approximations\cite{Recher_Sukhorukov_prl_2000,Wegewijs_Nazarov_arXiv_2001}, as explained in Appendix \ref{appendixB}). 
First, we assume  that the electrodes have a flat density of states within a bandwidth larger than all relevant energy scales in the problem: temperature, bias and the excitations energies of the MS within the $q_0$ manifold.   Second, we  neglect the energy dependence of the hopping matrix elements   $V_{k\eta,i}=V_{\eta,i}$.\cite{Merino_Gunnarsson_prb_2004}  These approximations are justified in IETS experiments where the temperature is at most a few Kelvins and the applied bias is bellow $50$mV. If we introduce the excitation energy associated to the transition between $n'$ and $n$ states in the $q_0$ manifold, $\Delta_{nn'}=E_n-E_{n'}$ and we define the average energy
$\bar{\epsilon}^{\eta\eta'}_{nn'}=1/2(\mu_{\eta}+\mu_{\eta'}+\Delta_{nn'})$,
 the transition rates $W_{nn'}^{\eta\eta'}$ obtained in Appendix \ref{appendixB} can be expressed as
\beqa
W_{nn'}^{\eta\eta'}\approx \sum_{\sigma\sigma'}\frac{2\pi\rho_{\eta\sigma} \rho_{\eta'\sigma'}}{\hbar}  
{\cal G}(\mu_{\eta}-\mu_{\eta'}+\Delta_{nn'}) 
 \Sigma^{\eta\sigma,\eta'\sigma'}_{n n'}(\bar{\epsilon}^{\eta\eta'}_{nn'})
 \crcr &&
\label{ratesapprox}
\eeqa
where ${\cal G}(\ome)=\frac{\omega}{1-\exp{\left[-\beta\omega\right]}}$ and
$\rho_{\eta\sigma}$  are the  spin and electrode resolved density of states. The many-body matrix elements $\Sigma^{\eta\sigma,\eta'\sigma'}_{n n'}\left(\bar{\epsilon}\right)$ are given by
\beqa
\Sigma^{\eta\sigma,\eta'\sigma'}_{n n'}(\bar{\epsilon})=
\left|\langle n | \left(
\hat{{\cal O}}^{(+)}_{\bar{k}\eta\sigma,\bar{k}'\eta'\sigma'}
 -\hat{{\cal O}}^{(-)}_{\bar{k}'\eta'\sigma',\bar{k}\eta\sigma}\right)|n' \rangle  \right|^2,
\label{Xidef}
\eeqa
where   $\bar{k}\equiv k(\bar{\epsilon})$, i.e., the quasiparticle energy that appear in the denominators are replaced by the corresponding 
bias-dependent average energy $\bar{\epsilon}_{nn'}^{\eta\eta'}$.

In this context, the current is given by\cite{Delgado_Rossier_prb_2010,Delgado_Palacios_prl_2010}
\begin{equation}
I_{T\rightarrow S}=e\sum_{n,n'} P_n(V) \left(W_{n,n'}^{S\rightarrow T}-W_{n,n'}^{T\rightarrow S} \right)
\label{current}
\end{equation}
where $e$ is the (negative) electron charge. 
This equation has  a physically transparent meaning: the current is proportional to the transition rates of quasiparticles  changing electrode. These rates involve transitions of the MS from the state $n$, which is occupied with probability $P_n(V)$, to state $n'$, including  elastic events $n=n'$.

Our convention  for the applied bias is such that $eV=\mu_S-\mu_T$ (electrons move from tip to surface for a positive applied bias). 
The bias implies  a small charge accumulation both in the tip and the surface which in turn involves a shift of their chemical potentials with respect to their equilibrium value, denoted by $E_F$. Without loss of generality we can write $\mu_S= E_F + x eV$, $\mu_T= E_F + (x-1)eV$, where $x$  is an undetermined  parameter that relates the bias voltage to the shift of the chemical potential in each electrode. Given the fact that the capacitance of the surface is much larger than that of the tip, it is reasonable to take $x=0$.  As we show below, this assumption makes it possible to account for the conductance asymmetry reported experimentally.\cite{Chen_Fu_prl_2008,Fu_Zhang_prl_2009}

In the following, we shall express the differential conductance in units of 
\beqa
g_0=\frac{G_0}{2}\rho_S\rho_T\left(J^2_{TS}+{\cal W}_{TS}\right),
\eeqa
 where $G_0=2e^2/h$ is the quantum of conductance, $\rho_\eta=\sum_{\sigma}\rho_{\eta\sigma}$ and $J_{TS}$ and ${\cal W}_{TS}$ are just the generalizations of the (momentum independent) exchange and direct coupling respectively that appears in the Anderson model, as it will be shown bellow:
\beqa
J_{TS}=2V_{S}^{(M)}V_{T}^{(M)} \left[(\mu_e-E_F)^{-1}+(E_F-\mu_h)^{-1}\right],
\label{Jgen}
\eeqa
and 
\beqa
{\cal W}_{TS}=V_{S}^{(M)}V_{T}^{(M)}/2\left[(\mu_e-E_F)^{-1}-(E_F-\mu_h)^{-1}\right],
\eeqa
where $V_{\eta}^{(M)}$ is the maximum value of the couplings between electrode $\eta$ and the orbitals of the MS.

\subsection{Summary of the method}
The approach described above can be implemented in a wide range of situations  following a sequence of well defined steps: 
\begin{enumerate}

\item Diagonalization of the MS Hamiltonian in  the 3 relevant charge sectors, $q=q_0-1,\;q_0,\;q_0 +1$, providing
 $|q,n\rangle$ and $E_{q,n}$.

 \item Computation of the  matrix elements (\ref{oplus}) and (\ref{ominus}) of the effective tunneling Hamiltonian operator, which  requires the calculation of the  many body matrix elements $\gamma$ (Eqs. (\ref{gplus}) and (\ref{gminus}) ) and the   ${\cal O}$-matrix prefactors.  
 
 \item Calculation of the scattering rates (\ref{ratesapprox}), which depend on bias, temperature,  MS-electrode coupling,  electrode density of states and MS wave functions. 
 
 \item Finding the non-equilibrium steady state solutions $P_n(V)$  of  the master equation (\ref{master}). 
 
 \item Evaluation of the current using Eq. (\ref{current}).

  \end{enumerate}

\subsection{Comparison with other cotunneling theories}
The calculation of cotunneling current has been widely studied before, using different methodologies, mainly in the context of quantum dots\cite{Averin_Nazarov_prl_1990,Averin_Nazarov_book,Beenakker_prb_1991} and, more recently, 
molecules.\cite{Elste_Timm_prb_2005,Elste_Timm_prb_2007,Hansen_Mujica_nanolett_2008,Roch_Vincent_arXiv_2011,Tikhodeev_Ueba_book_2008}
  For instance, in Ref.~\onlinecite{Elste_Timm_prb_2005,Elste_Timm_prb_2007,Hansen_Mujica_nanolett_2008,Roch_Vincent_arXiv_2011} they compute the cotunneling scattering rates by truncating the T-matrix down to second order in the electrode coupling.
On the other hand, a more formal and accurate treatment, valid also in the strong-coupling regime, was introduced in Ref.~\onlinecite{Tikhodeev_Ueba_book_2008}, where the non-equilibrium Keldysh Green function formalism was used to study the inelastic spectroscopy of single adsorbed molecules. 

Whereas there current obtained  using these different methods is the same, our approach permits to derive an effective Hamiltonian which, in the adequate limit,  is the same than the effective Kondo Hamiltonian used extensively in previous works.\cite{Appelbaum_pr_1967,Rossier_prl_2009,Fransson_nanolett_2009,Persson_prl_2009,Delgado_Palacios_prl_2010,Fransson_Eriksson_prb_2010,Zitko_Pruschke_njphys_2010,Sothmann_Konig_njp_2010,Delgado_Rossier_prb_2010} 
An interesting work addressing the relation between multiple-impurity Anderson model at half-filling and a Kondo model was presented
in Ref.~\onlinecite{Zitko_Bonca_prb_2006}, where authors proved that a Hubbard chain of $N$ impurities coupled in parallel can be described with a $S=N/2$ $SU(2)$ spin Kondo model. In Ref.~\onlinecite{Bas_Aligia_jphys_2010}, authors used the same generalized Schrieffer-Wolff transformation to relate a singlet-triplet Anderson impurity with a spin model close to its quantum phase transition.

Our approach, based on an effective cotunneling Hamiltonian directly obtained from the {\em exact} description of the magnetic system, provides a microscopic justification of earlier phenomenological works, at the time that it keeps the simplicity that allows to calculate the current  as described above. 

%


\section{Single orbital Anderson model\label{Anderson}}
In this section we revisit the  very well known Anderson model\cite{Anderson_prl_1966} for which the MS is a single site Hubbard model: 
\beqa
{\cal H}_{\rm MS}=E_d \sum_\sigma d^\dag_{\sigma} d_\sigma+U n_\downarrow n_\uparrow,
\label{handerson}
\eeqa
where $E_d$ is the on-site energy level, $U$ the on-site Coulomb repulsion and $n_{i\sigma}=d_{i\sigma}^\dag d_{i\sigma}$.
We now derive an effective cotunneling Hamiltonian which, as we show below, turns out to be identical to the spin 1/2 Kondo model by means of a Schrieffer-Wolff transformation.\cite{Schrieffer_Wolff_pr_1966,Anderson_prl_1966}  By so doing, we test the validity of our approach and shed some light on the origin of the large contribution of the
inelastic spin assisted tunneling to the conductance. 

The single-site Hubbard Hamiltonian has only 3 possible charge states, empty, singly and doubly occupied.
The singly occupied  manifold has two states, $|\uparrow\rangle$ and $|\downarrow\rangle$ with energy $E_d$. 
The empty and doubly occupied manifolds have only 1 state each, $|\uparrow\downarrow\rangle$ with energy $2E_d + U$  for the $+$ manifold and  $|0\rangle$, with energy $0$  for the $-$ manifold respectively.  If $E_d+U \gg E_F \gg E_d\gg k_bT$ the ground state has  $q_0=1$ and classical charge fluctuations are frozen. 
Hence, the virtual transition operators acting on the $q_0=1$ space  have dimension two and can be expressed as Pauli matrices, acting on the spin space. 

After a straightforward calculation we find the effective cotunneling Hamiltonian with 3 contributions.
First, the famous exchange assisted Kondo term:\cite{Schrieffer_Wolff_pr_1966,Anderson_prl_1966}
\beqa
{\cal H}_{\rm cot,1}= \sum_{kk',\eta\eta',\sigma\sigma'} 
J_{kk',\eta\eta'}\vec{{\cal S}}.\vec{\tau}_{\sigma\sigma'} f_{k\eta\sigma}^\dag f_{k'\eta'\sigma'},
\eeqa
with
\beqa
J_{kk',\eta\eta'}=V_{k\eta}V_{k'\eta'}^*\left[\frac{1}{E_d+U-\epsilon_{k\eta\sigma}}
+\frac{1}{\epsilon_{k'\eta'\sigma'}-E_d}\right].
\label{JAnd}
\eeqa
The second term ${\cal H}_2=\sum_{kk',\eta\eta',\sigma\sigma'}{\cal H}_2(kk',\eta\eta',\sigma)$ in the Hamiltonian corresponds to a direct (spin-independent) interaction, also obtained in the Schrieffer-Wolff transformation\cite{Anderson_prl_1966}

\beqa
{\cal H}_{\rm cot,2}= \sum_{kk',\eta\eta',\sigma}
 {\cal W}_{kk',\eta\eta'}  f_{k\eta\sigma}^\dag f_{k'\eta'\sigma},
\eeqa
where 
\beqa
{\cal W}_{kk',\eta\eta'}=V_{k\eta}V_{k'\eta'}^*\left[\frac{1}{E_d+U-\epsilon_{k\eta\sigma}}
-\frac{1}{\epsilon_{k'\eta'\sigma'}-E_d}\right]
\eeqa

Notice how in this model, the exchange assisted $J_{kk',\eta\eta'}$ and the direct tunneling term ${\cal W}_{kk',\eta\eta'}$ have a common origin, namely, virtual charging of the  magnetic site.  Importantly, we see how we can have the spin-flip term much larger than the direct term. In particular, in the so called symmetric case, for which $E_d+U-E_F = E_F-E_d$, the direct term vanishes altogether, due to a cancellation between the electron addition and hole addition channels.  In that situation only the spin-flip assisted tunneling would be possible.  Thus, the cotunneling picture provides a natural scenario for the large contribution of the inelastic contribution to the conductance.    
Finally, a third term ${\cal H}_3=\sum_{k,\eta,\sigma}{\cal H}_3(k,\eta,\sigma)$ is obtained, which can be 
considered as a renormalization of the on-site energy level.

\section{Stacks of ${\rm CoPc}$ molecules\label{molecules}}
In this section we model the  IETS experiments of stacks of Cobalt phthalocyanine molecules (CoPc)
 deposited on Pb$(111)$.\cite{Chen_Fu_prl_2008,Fu_Zhang_prl_2009}  
 CoPc molecules are planar molecules with $D_{4h}$ symmetry and a single Cobalt (Co) atom at its center, surrounded by four Nitrogen neighbors and enclosed by aromatic macrocycles.  A single CoPc has a ground state with spin $S=1/2$, corresponding to an unpaired electron presumably in the $d_{z^2-r^2}$ orbital of Co. 
 In a stack with $N+1$ CoPc  molecules, the  CoPc in contact with the Pb surface  acts as a dead layer that isolates the remaining $N$ molecules.
 
  The stacking seems to be such that Cobalt atoms are underneath Nitrogen atoms of the adjacent molecule. 
   The IETS results\cite{Chen_Fu_prl_2008,Fu_Zhang_prl_2009} of stacks  with $N$ active CoPc (N+1 molecules in total) can be interpreted as if the molecules are coupled via an antiferromagnetic coupling,
   which presumably comes from super-exchange between two Cobalt  coupled to a common Nitrogen.
   The observed spin-flip
excitations were successfully described using a Heisenberg model with an antiferromagnetic (AF) coupling $J\simeq 18$meV .    

Whereas the Heisenberg model accounts for the observed excitation energies, it can not account for either the transport mechanism or the  fact that the conductance in this system is very asymmetric.  In particular, some inelastic steps  seen at a given bias polarity are not seen when bias sign is reversed.
 Additional    experiments where the charge state of the molecular stack was controlled using the STM tip as a local gate make it necessary to go beyond spin-only models.\cite{Fu_Zhang_prl_2009}  The observed excitation energies could be accounted for using a Hubbard model, rather than a Heisenberg model:
\beqa
{\cal H}_{\rm MS}=E_d\sum_{i\sigma}d_{i\sigma}^\dag d_{i\sigma}+U\sum_{i}n_{i\downarrow}n_{i\uparrow}\nonumber\\
+ t \sum_{i,\sigma} \left(d_{i\sigma}^\dag d_{i+1\sigma} +{\rm h.c.}\right)
\label{Hhubb}
\eeqa
Here $E_d$ stands for the energy of the $d_{z^2-r^2}$ orbital with respect to the Fermi energy, that we take at $0$, $U$ stands for the on-site Coulomb repulsion and $t$ for the Co-Co hopping, which  actually occurs through the common Nitrogen neighbor.  
   In the strongly insulating limit, $U\gg t$ and at half-filling (1 unpaired electron per Cobalt atom),  the Hubbard model    
    has the same low energy excitation spectra than the Heisenberg model with $J=\frac{4t^2}{U}$. 
    Away from half-filling, when  the molecular stack is charged, the mapping to the Heisenberg model is no longer possible but still the excitation energies observed experimentally are accounted for by the Hubbard model.\cite{Fu_Zhang_prl_2009} 
Here we focus on the half-filling case and we apply our formalism to short Hubbard chains with $N=2,3,4$ sites.  We  take $U=1.5$eV which imposes  $t= 82$meV ($J\approx 18$meV), in accordance with the experimentally observed value.\cite{Chen_Fu_prl_2008,Fu_Zhang_prl_2009}


\subsection{The dimer\label{dimer}}
 
\begin{figure}
\begin{center}
\includegraphics[angle=0,width=0.9\linewidth]{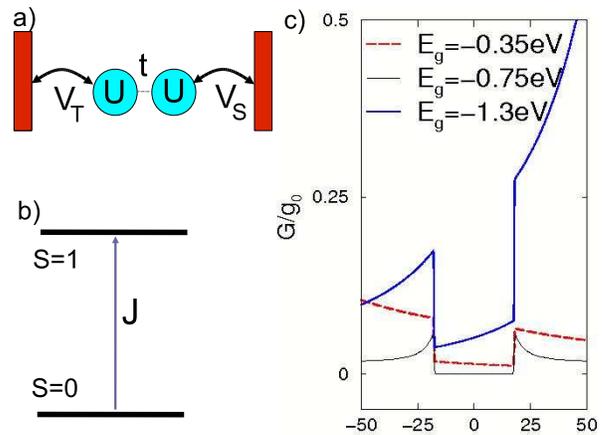}
\end{center}
\caption{ \label{fig2n}(Color online) (a) Scheme of a two sites Hubbard dimer connected to two electrodes. (b) Lowest energy levels of Hubbard dimer at half filling in terms of the exchange constant $J$. $dI/dV$ as a function of applied bias. Here $\rho_S V_S=5$ and $\rho_T  V_T=1$ and $T=0.4$K.
}
\end{figure}

 The  eigenvalues and eigenvectors of the Hubbard dimer 
  can be found analytically, both for the half-filling sector and the two sectors with $1$ and $3$ electrons. 
At half filling $(q=q_0)$ the ground state  corresponds to a spin singlet, $S=0$, while the first excited state corresponds to a spin triplet, $S=1$ with excitation energy $J$ neglecting terms of order $t^4/U^3 $, see Fig.~\ref{fig2n}(b) (in agreement with the  experimental results\cite{Chen_Fu_prl_2008,Fu_Zhang_prl_2009}).
Referred to  half-filling, the electron addition and hole addition energies are $E_d+ U$  and
$-E_d$, respectively.   Thus,  for the MS to be at half filling we must have $E_d<0$ and $|E_d|<U$. 
  The states of the $q=\pm$ sectors correspond to those of a single electron and a single hole, respectively.  

In order to assess the effect of the relative weight of the cotunneling mediated by virtual hole and virtual electron addition ($q_{\pm}$ channels), 
 we 
have  calculated $dI/dV$ spectra obtained for the  Hubbard  dimer  for three different values of the on-site energy $E_d$: $E_d= -U/2$, the so called electron hole symmetry point, $E_d= -0.35$eV for which virtual transitions to the $q_-$ manifold  are favored  and $E_d= -1.3$eV, which favors  virtual transitions to the $q_+$ manifold, see Fig.~\ref{fig2}(c).  Whereas
 the excitation step at $\pm eV=J$  is present in all of them,  both the magnitude and the bias dependence of the elastic contribution depends a lot on $E_d$.
 At the electron-hole symmetry point (EHSP), the elastic conductance is zero, as in the Anderson model, and
 the non-monotonic lineshape right above the inelastic step is due to the depletion of the occupation of the ground state in favor of the excited state, a non-equilibrium effect  
 discussed in our previous work.\cite{Delgado_Rossier_prb_2010}
 Both the elastic and the inelastic contributions increase when $E_d$ is taken away from the EHSP.

When $E_d=-1.3$eV, the virtual transition to the $q_+$  manifold is dominant and  cotunneling is mediated by the addition of an electron. As we mentioned in Sec.~\ref{rates}, our bias convention is such that positive bias $V$ results in an increment of the tip chemical potential with respect to the molecules and the surface.   Thus, for $V>0$  it becomes easier to add an electron to the system, increasing the global conductance. For $V<0$, instead, the chemical potential of the tip is decreased, making it relatively harder to  charge the dimer with an electron and reducing the cotunneling conductance thereby.  
In the  case of $E_d=-0.35$eV  the situation is  reversed. The virtual transition to the $q_-$ manifold is dominant, ie., cotunneling is mediated by the addition of a hole (or the removal of an  electron).  In this case a positive bias makes it harder for the electron to tunnel out of the system, decreasing the conductance. 
Thus, in our calculation the asymmetry of the conductance comes from the assumption that the bias shifts mostly the tip chemical potential, and not the surface, and the fact that one of the two cotunneling channels (virtual addition of either an electron or a hole) is dominant. 
Comparing with the experimental results,\cite{Chen_Fu_prl_2008} we infer that the double CoPc molecule system is  close to  the electron addition  point. This has been further confirmed by additional experiments
  by the same group.\cite{Fu_Zhang_prl_2009}

\subsection{The trimer and the tetramer}

\begin{figure}[t]
\begin{center}
\includegraphics[angle=0,width=0.9\linewidth]{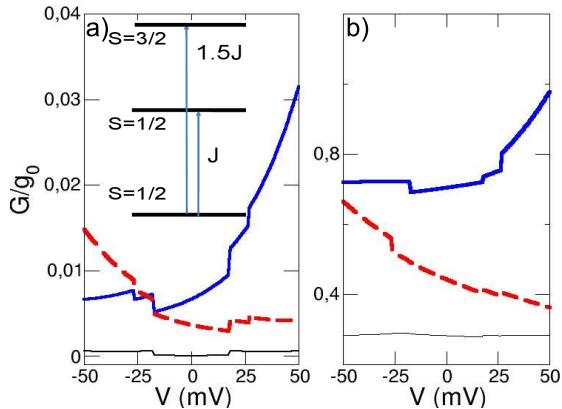}
\end{center}
\caption{ \label{trimer} (Color online) $dI/dV$ as a function of applied bias for $N=3$ with on-site energy $E_d=-0.35$eV (red-dashed line), $E_d=-0.75$eV (thin-black line) and $E_d=-1.2$eV (thick-blue line). (a) Serial Hubbard trimer with $V_{S,i}=2\delta_{i,3}$ and $V_{T,i}=5\delta_{i,1}$. (b) Multiple electrode connected Hubbard trimer with 
 with $\rho_{S}{\bf V_S}=(2,1,1)$ and $\rho_{S}{\bf V_S}=(2,3,5)$. Inset: scheme of the lowest energy levels. The other parameters are kept as in Fig.~\ref{fig2}.
}
\end{figure}

\begin{figure}[t]
\begin{center}
\includegraphics[angle=0,width=0.9\linewidth]{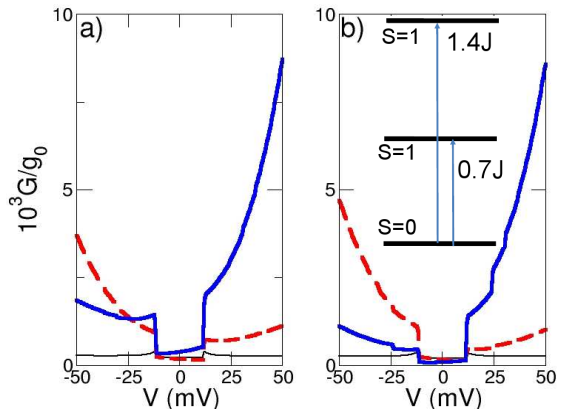}
\end{center}
\caption{ \label{tetramer} (Color online) $dI/dV$ as a function of applied bias for $N=4$ with on-site energy $E_d=-0.35$eV (red-dashed line), $E_d=-0.75$eV (thin-black line) and $E_d=-1.2$eV (thick-blue line). (a) Serial Hubbard tetramer with $V_{S,i}=3\delta_{i,4}$ and $V_{T,i}=10\delta_{i,1}$. (b) Multiple electrode connected Hubbard tetramer with 
 with  $\rho_{S}{\bf V_S}=(0,0,0.2,10)$ and $\rho_{t}{\bf V_S}=(3,0.3,0,0)$. Inset: scheme of the lowest energy levels. The other parameters are kept as in Fig.~\ref{fig2}.
}
\end{figure}

We now consider the Hubbard chains  with either $N=3$ and $N=4$ sites
  and try to model the CoPc molecular stacks with 3 and 4 active molecules respectively.\cite{Chen_Fu_prl_2008}
We assume that $t$ and $U$ take the same values than before and that there is one electron per site in the ground state.
   We label the sites from $n=1$ to $n=N$, starting from the molecule closest to the tip.
For $N=3$, the ground state and first excited state have $S=1/2$ and the second excited state has $S=3/2$ (see Fig.~\ref{trimer}).  Thus, we expect two inelastic transitions, at energies  $J$ and $3J/2$.   For the $N=4$ chain, the ground state has $S=0$ and the two lowest energy excited states, both with $S=1$, have excitation energies $0.7J$ and $1.4 J$, see Fig.~\ref{tetramer}.   Again, two inelastic steps are expected at those energies. 

In Figs.~\ref{trimer}(a) and \ref{tetramer}(a) we show the conductance for $N=3$ and $N=4$ respectively assuming that the electrons can tunnel from the tip to the $n=1$ site only and from the $n=N$ site to the surface only.  As in the case of the dimer, we take 3 different values for $E_d$:   hole mediated, electron mediated and EHSP. On top of the symmetry trends already discussed for the dimer,  we see how in the EHSP only the lowest energy transition is seen both in the $N=3$ and $N=4$ cases.  This suggests that not only the elastic contribution vanishes, as in the case of the Anderson model and the Hubbard dimer, but also some of the  inelastic transitions can be suppressed possibly due to the destructive interference between the hole and electron channels.

 In the case of Figs.~\ref{trimer}(a) and \ref{tetramer}(a), where only the sites at the end of the chain are coupled to either the tip or the surface,  the steps are visible for both signs of $V$, at odds with the experimental observations.\cite{Chen_Fu_prl_2008}  In an attempt to explore a scenario in which the height of the steps are only visible at a given polarity, we have considered a situation where electrons can tunnel from the tip to sites other than $n=1$ and from the surface to sites other than $n=N$.  By so doing, we can obtain $dI/dV$ curves where the steps are depleted  for $V<0$ (Figs.~\ref{trimer}(b) and \ref{tetramer}(b)). However, we think that a more  plausible explanation would come from a microscopic calculation including more than 1 orbital per molecule.
 It must also be mentioned the broadening of the excitations observed  experimentally is larger than 5.4 $k_BT$, which indicates 
than neglecting the intrinsic broadening due to the coupling to the continuum of states of the electrodes is not fully justified.\cite{Datta_book_2005}

%
%
%

\section{Magnetic adatoms\label{chain}}
We now consider spin IETS through a single Mn atom deposited on a Cu$_2$N surface. This system has been widely studied experimentally \cite{Hirjibehedin_Lutz_Science_2006,Hirjibehedin_Lin_Science_2007,Loth_Bergmann_natphys_2010}
and
 theoretically,\cite{Rossier_prl_2009,Fransson_nanolett_2009,Persson_prl_2009,Delgado_Palacios_prl_2010,Fransson_Eriksson_prb_2010,Zitko_Pruschke_njphys_2010,Sothmann_Konig_njp_2010,Delgado_Rossier_prb_2010,Gauyacq_Novaes_arXiv_2010,Novaes_Lorente_prb_2010}  in most instances modeling  the Mn spin with an effective spin model.  Here we go beyond the spin model picture and we use a multiorbital Anderson Hamiltonian for the  5 $d$ electrons of the Mn$^{+2}$ ion which includes Coulomb interaction, spin-orbit coupling and crystal field.  Transport occurs via virtual transitions to the many-body states with either 4 or 6 $d$ electrons.  Our approach requires the 
exact diagonalization of the fermionic model in the 3 relevant charge states, with 4, 5 and 6 electrons.  Below we  describe the multi-orbital Anderson model, the transport calculation and compare with  the experimental results in Ref.~\onlinecite{Hirjibehedin_Lin_Science_2007}.

\subsection{Magnetic system Hamiltonian\label{CImethod}}
Here we describe our  model Hamiltonian for the Mn ion in the Cu$_2$N surface.  The purpose of our model is to provide a minimal fermionic Hamiltonian that accounts for the data, rather than to provide a realistic description of the Mn ion on the surface. Density functional calculations\cite{Hirjibehedin_Lin_Science_2007,Lorente_Gauyacq_prl_2009,Rudenko_Mazurenko_prb_2009} suggest that the Mn adatom transfers charge to the CuN surface
and creates bonds with its neighboring N atoms. As a result, the Mn adatom becomes a Mn$^{2+}$ ion that has lost its two $4s$ electrons. 
 We  model this system considering only the $3d^5$ electrons of the Mn, including  the electrostatic potential of the  neighboring atoms. The Hamiltonian of the MS  can be 
 written as:
 \beqa
{\cal H}_C=H_{ee}+H_{CF}+H_{SO}+H_{Zeem},
\label{Htot}
\eeqa
where $H_{ee}$ is the Coulomb repulsion between the $3d$ electrons, $H_{CF} $ is the crystal field Hamiltonian, $H_{SO}$ is the 
spin-orbit Hamiltonian and $H_{Zeem}$ is the Zeeman Hamiltonian associated to an applied magnetic field $\vec{B}$. 
The Coulomb matrix elements  of the atomic orbitals can be expressed in terms of radial integrals which depend on the specific form of the approximate wave function and an angular part that can be obtained analytically.\cite{Racah_pr_1942}
We have taken them from a calculation for an isolated ion using Gaussian package\cite{gausssian_09} which yields to the unscreened on-site Coulomb repulsion $U\simeq 24$eV, in accordance with unscreened Hartree-Fock calculations\cite{Schnell_Czycholl_prb_2003}.  Since screening in the real system makes $U$ much smaller than the single ion calculation we have downscaled 
the Coulomb matrix elements with an overall  dielectric constant of $\varepsilon=4.7$ in order to obtain $U$ in the range of 5eV.\cite{Jacob_Haule_prl_2009}

 The energy  $E_d$ of the $d$ levels before crystal splitting is included,  is kept as a free parameter in our theory. The crystal field term $H_{CF}$,  is built using  a point charge model for the first N and Cu neighbors,\cite{Dagotto_book_2003}  whereas an effective dielectric constant $\varepsilon'$ 
was introduced to account for the screening of the bare crystal field and fit the many-body spectrum to that of the single ion Hamiltonian.\cite{Zhao_Chiu_prb_1995} 
Fig.~\ref{fig5}(a) shows the splitting of  the five $d^5$ energy levels due to the crystal field, together with its dominant orbital contribution. 
Finally, the spin-orbit Hamiltonian reads:
\beqa
H_{SO} = \frac{\lambda}{\varepsilon''}\sum_{m,m',\sigma,\sigma'} \langle m\sigma|\vec{L}\cdot\vec{S}|m'\sigma'\rangle
d^{\dagger}_{m\sigma}d_{m',\sigma'}
\label{HSO}
\eeqa
where $\lambda=43$meV corresponds to the value of the bare Mn$^{2+}$ ion\cite{Zhao_Chiu_prb_1995}  and $\varepsilon''$ is another free parameter in our model. 

The Hamiltonian (\ref{Htot}) corresponding to the $3d^5$ electrons was then diagonalized in the space of the 252 possible configurations, using the configuration interaction (CI) method. Analogously, the eigenvalues and eigenvectors of the $3d^4$ and $3d^6$ configurations were calculated in order to get the transition rates (\ref{gamma}). 
The condition of stable configuration with $N_e=5$ electrons require that $E_G(5)\le E_G(4),E_G(6)$. In our case, this bound translates into the inequality $-24.1{\rm eV}<E_d<-18.9{\rm eV}$. In particular, we choose $E_d$ in the middle of this energy window and, as it will be shown in next section, results do not change significantly with $E_d$.

\subsection{Mn$^{2+}$ energy spectra\label{resultsci}}
According to first Hund's rule, we expect that  the spin of the ground state for the  half filled $d$ shell is $S=5/2$, which is what we obtain from the diagonalization of the model.   The sixfold degeneracy  at zero field is broken by the combined action of 
spin-orbit and crystal field.   
Due to the spin-orbit coupling, the total spin $S$ and total angular momentum $L$ are no longer good quantum numbers. However, our CI method allows to calculate any of these expectations values. We have verified  that for
 our calculation for the Mn$^{2+}$, $\langle S\rangle \approx 5/2$, while $\langle L\rangle\approx 0$, with a deviation smaller than 0.1$\%$. In the same way, $S_z$ is almost a good quantum number.

\begin{figure}[t]
\begin{center}
\includegraphics[height=1.\linewidth,width=0.7\linewidth,angle=-90]{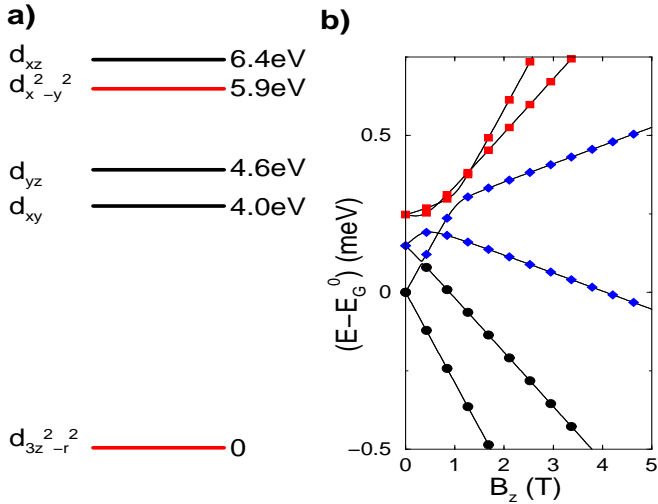}
\end{center}
\caption{ \label{fig5} (Color online) (a) Single particle energy levels of the crystal field Hamiltonian with the dominant orbital contribution (only red levels are coupled by the crystal field).(b) Lowest energy spectra of the total Hamiltonian ${\cal H}_{MS}$ for the Mn$^{2+}$ ion on a Cu$_2$N surface as studied in Ref. \onlinecite{Hirjibehedin_Lin_Science_2007}. The dots symbols
corresponds to the solution of the phenomenological spin model with $D=-0.39$meV and $E=0.06$meV.
Here $E_{d}=-21.5$eV,  $\varepsilon'=11.3$ and $\varepsilon''=1.9$.}
\end{figure}

The location of the first neighbors of Mn is taken from reference \onlinecite{ Hirjibehedin_Lin_Science_2007}. The values of  $\varepsilon'$ and $\varepsilon''$ are taken so that the lowest energy levels of the  energy spectra obtained from the diagonalization of (\ref{Htot}) are in agreement with those of the single ion Hamiltonian, as shown in the Fig.\ref{fig5}(b).  
At zero field, the lowest energy doublet  corresponds to $\langle S_z\rangle \approx \pm 5/2$. For the 
two pairs of excited levels, we get $\langle S_z\rangle \approx \pm 3/2$ and $\langle S_z\rangle \approx \pm 1/2$, in order of increasing energy.
Fig.~\ref{fig5}(b) shows the magnetic field dependence of the low energy spectra of the Mn$^{2+}$ obtained using the CI calculation, together with the fitting to a phenomenological spin model\cite{Heinrich_Gupta_science_2004,Hirjibehedin_Lutz_Science_2006,Hirjibehedin_Lin_Science_2007,Otte_Ternes_natphys_2008,Loth_Bergmann_natphys_2010,Tsukahara_Noto_prl_2009,Zhou_Wiebe_natphys_2010}
\beqa
{\cal H}_{S}=DS_z^2+E(S_x^2-S_y^2)+g\mu_B \vec{B}.\vec{S}.
\label{hspin}
\eeqa
The first two terms in Eq. (\ref{hspin}) describe the single ion magneto-crystalline anisotropy  while the last one
corresponds to the Zeeman splitting term under an applied magnetic field $\vec{B}$. The main magnetization direction {\em z} in Eq. (\ref{hspin}) depends on the substrate and magnetic atom nature. In the case of the Mn on a Cu$_2$N substrate, the $z$-axis is perpendicular to the surface. This result is also reproduced by our model (\ref{Htot}).   

\subsection{Transport\label{transportci}}
Once the many-body eigenstates of  Hamiltonian (\ref{Htot}) are obtained, we are in position to study transport through the magnetic atom.  For that matter, we need to specify the coupling of the 5 $d$ orbitals to the tip and the substrate.
As the  $dI/dV$ spectra was recorded with the tip located exactly over the Mn atom, we will 
 assume that the tip-atom tunneling is dominated by tunneling between the tip apex $s$ orbital and the $d_{3z^2-r^2}$,\cite{Slater_Koster_pr_1954} oriented along the adatom-tip axis.  For the coupling with the substrate, the situation is significantly more complicated and we 
 couple equally all the $d$-orbitals to the substrate, i.e., $V_{S,i}=V_S$. 
For simplicity, we will omit the coupling between the $s$-orbital of the Cu atom and the empty $s$-orbital of the Mn$^{2+}$ ion at odds with existing DFT calculation.\cite{Novaes_Lorente_prb_2010} Another important parameter to properly account for the transport properties of the system is the Fermi level of the electrodes. Here we have assumed that the Fermi level of the $Cu$ substrate 
coincides with its bulk Fermi level, $E_F=-7$eV.\cite{Ashcroft_Mermin_book_1976}

\begin{figure}[t]
\begin{center}
\includegraphics[height=1.\linewidth,width=0.65\linewidth,angle=-90]{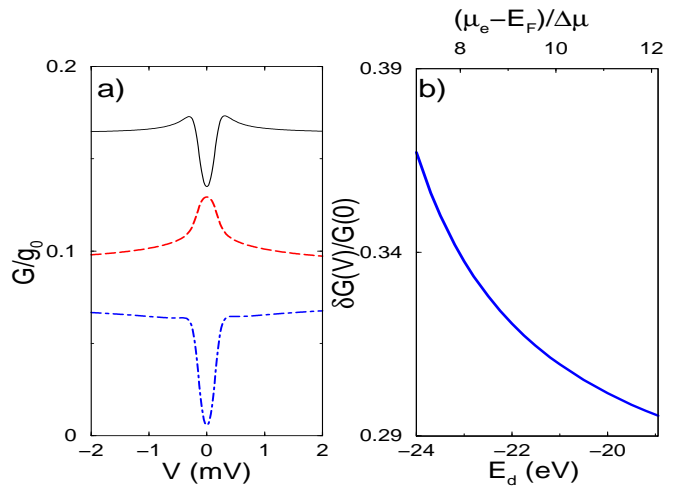}
\end{center}
\caption{ \label{fig6} (Color online) (a) Total (solid line), elastic (dashed line) and inelastic (dotted-dashed line) $dI/dV$ as a function of applied bias for the Mn$^{2+}$ ion ($\rho_S V_{S}=1$, $\rho_T V_{T,3z^2-r^2}=1$, $T=0.4$K and $E_{d}=-21.5$eV).
(b) Ratio $G^{ine}(V)/G^{elas}(0)$ for $V=2$meV versus on-site energy $E_d$ (lower axis) or $(\mu_e-E_F)/\Delta \mu$ (upper axis), with $\Delta \mu=\mu_e-\mu_h$. Other parameters as in Fig.~\ref{fig5}.}
\end{figure}

The resulting $dI/dV$ is plotted in Fig.~\ref{fig6}(a), were the elastic, inelastic and total differential conductance are plotted. Our calculation reproduces both the  line-shape of the $dI/dV$ curves as well as the  the relative contribution between the elastic and inelastic parts, $G_{\rm inel}/G_{\rm el}\simeq 0.5$ .\cite{Hirjibehedin_Lin_Science_2007,Loth_Bergmann_natphys_2010}  Within the model this ratio depends on the position of the charging energies of the atom, $\mu_e$ and $\mu_h$, with respect to the chemical potential of the electrodes.   
In Fig.~\ref{fig6}(b) we show  ratio $G_{\rm inel}/G_{\rm el}$ as a function of the  on-site energy level $E_d$ in
the window of energies where the system ground state contains 5 electrons.
 As observed, the ratio $G_{\rm inel}/G_{\rm el}$ varies smoothly between $0.4$ and $0.6$.  Thus, our model yields a large inelastic signal, consistent with the experiments,  without fine tuning the on-site energy $E_d$.  Notice that  in the case of Mn on Cu$_2$N at $T=0.4$K, the thermal broadening of the inelastic step is such that the inelastic conductance is non-zero even at zero bias. 

\section{Discussion and Conclusions\label{conclusions}}
We have shown  that the inelastic tunneling spectroscopy widely used to study magnetic molecules and atoms adsorbed on surfaces, can be understood in terms of cotunneling.\cite{Averin_Nazarov_prl_1990,Averin_Nazarov_book} As  the electrons go from the tip to the surface, the magnetic system must undergo virtual transitions to states with an extra electron or an extra hole. This picture holds both for elastic tunneling, in which case  the MS  returns to the original state after the virtual charging process, and inelastic tunnel, for which the state before and after the virtual charging are different.  Thus, the origin of both elastic and inelastic conductance is the same, which accounts for the large inelastic signal reported  experimentally in a variety of systems, including Mn, Fe and Co on Cu$_2$N\cite{Hirjibehedin_Lin_Science_2007} or Fe on InSb.\cite{Khajetoorians_Chilian_nature_2010}
Further support to this claim comes from comparison of the evolution of the $dI/dV$ as a function of an applied magnetic field
of a quantum dot with a single resident  electron in the Coulomb Blockade regime\cite{Kogan_Amasha_prl_2004}
 and a single Cobalt atom on Cu$_2$N, both 
undergoing a transition from the Kondo regime at low field to inelastic steps at large field.  Both systems show very similar $dI/dV$. In addition, our microscopic theory provides a natural starting point to describe both the appearance of Kondo correlations, and their  relation to the inelastic spin flips in the context of magnetic adatoms and molecules.

Our approach is  based on  the derivation of an effective cotunneling Hamiltonian acting only in the space of neutral configurations of the MS. The calculation of the effective Hamiltonian requires the exact diagonalization of the MS in the neutral subspace as well as the subspaces with one extra electron and one extra hole. From the formal point of view our results are in  agreement with previous works based on a  truncation of the $T$-matrix to second order in the coupling Hamiltonian. \cite{Elste_Timm_prb_2005,Elste_Timm_prb_2007,Hansen_Mujica_nanolett_2008,Roch_Vincent_arXiv_2011}  Our approach permits to obtain an effective cotunneling Hamiltonian that can be compared
with effective Kondo-like Hamiltonians proposed in most theoretical analysis of IETS
 experiments.\cite{Appelbaum_pr_1967,Rossier_prl_2009,Fransson_nanolett_2009,Persson_prl_2009,Delgado_Palacios_prl_2010,Fransson_Eriksson_prb_2010,Zitko_Pruschke_njphys_2010,Sothmann_Konig_njp_2010,Delgado_Rossier_prb_2010}

We have also explored the origin of the experimentally observed asymmetry with respect to bias inversion in the $dI/dV$ curves.\cite{Hirjibehedin_Lutz_Science_2006,Hirjibehedin_Lin_Science_2007,Loth_Bergmann_natphys_2010,Loth_Lutz_njphys_2010,Chen_Fu_prl_2008,Tsukahara_Noto_prl_2009,Fu_Zhang_prl_2009}  It comes from a combination of two ingredients. First, we need to consider that the bias voltage results in a shift of the chemical potential in the tip, the one in the surface remaining constant. Second, the energy level alignment of the MS must be such that one of the  cotunneling  channels, either  virtual electron addition or virtual hole addition, is dominant.  

In summary, we propose a method to describe single spin inelastic electron tunneling spectroscopy which does not rely on effective spin models to describe both the magnetic system and the spin-flip assisted tunneling. Our approach provides a natural explanation for the large inelastic signals observed experimentally, and a microscopic mechanism for the spin assisted tunneling.

\begin{center}{\bf ACKNOWLEDGMENT }\end{center}

We acknowledge fruitful discussions with N. Lorente, C. F. Hirjibehedin, J. J. Palacios and C. Untiedt.  This work was supported by MEC-Spain (MAT07-67845,  FIS2010-21883-C02-01, Grants  JCI-2008-01885 and  CONSOLIDER CSD2007-00010) and Generalitat Valenciana (ACOMP/2010/070).



%
%
\appendix
\section{Effective tunneling Hamiltonian\label{appendixA}}
We now use Eq. (\ref{pert}) to derive an effective Hamiltonian which acts on the reservoir fermions and on the $q_0$ subspace of the central island only.  By so doing, we shall eliminate the $d^{\dagger}$ and $d$ operators from the effective Hamiltonian and, more important, we shall obtain a tunneling Hamiltonian for which the current can be derived straightforwardly. The matrix element between any two states in the $q_0$ manifold can be written as:
\begin{eqnarray}
\langle N|\hat{{\cal V}}_{\rm tun}|N'\rangle&=&
\langle \Psi_f(0)| \langle n |\hat{{\cal V}}_{\rm tun}|
\Psi_{f'}(0)\rangle |n'\rangle,
\end{eqnarray}
where $|N\rangle \equiv |n\rangle\otimes |\Psi_f(0)\rangle$, with $|\Psi_f\rangle$ a multi-electronic Slater state describing independent Fermi seas of left and right electrodes. Importantly, the unperturbed states are product states of the left and right electrodes and the central island. 
These states can describe both, the ground state of the MS with no excitations in the electrodes and excited states with an electron-hole pair in the electrodes and a excited state $n'$ in the central island. Notice that the electron-hole pair can be either in one electrode or split in the left and right electrodes. In the second case, this excitation contributes to the net current flow.
Now we need to evaluate matrix elements like
\begin{eqnarray}
\langle \Psi_f(0)| \langle n |{\cal V}^+ |M_-\rangle
&=&\sum_{\alpha,{\bf i}} V_{\alpha,{\bf i}}^*
\langle \Psi_f(0)|f_{\alpha}|\Psi_{mf}(-)\rangle
\crcr
&&\quad\times \langle n |d^{\dagger}_{{\bf i}}|m_-\rangle.
\label{element1}
\end{eqnarray}

Before going further, it is convenient to write down the explicit form of the electrodes wavefunctions. If we denote the ground state of the electrodes in the Fermi sea with no excitations and in its neutral charge state as $|0\rangle$, we can write $|\Psi_{f}(0)\rangle \equiv f_{\alpha}^\dag f_\alpha|0\rangle$, where we are creating an electron-hole pair with quantum number $\alpha$. For the states with one electron excess (defect) we will have $|\Psi_{mf}(-)\rangle= f_{\beta}^\dag f_{\alpha}^\dag f_\alpha|0\rangle$ ($|\Psi_{mf}(+)\rangle= f_{\beta} f_{\alpha}^\dag f_\alpha|0\rangle$). 
The matrix element of the electrode operator in Eq. (\ref{element1}) selects one and only one term in electrode part of the sums $\sum_{M-}=\sum_{m-}\sum_{mf}$.  The term in question is such that
\begin{eqnarray}
|\Psi_{mf}(-)\rangle= f^{\dagger}_{\gamma}| \Psi_f(0)\rangle.
\end{eqnarray}
This relation is equivalent to write $\langle \Psi_f(0)|f_{\gamma}|\Psi_{mf}(-)\rangle=(1-n_f(\gamma))\delta_{\beta\gamma}$, where $n_f(\gamma)=\langle \psi_f(0)| f^{\dagger}_{\gamma}f_{\gamma}|\psi_{f}(0)\rangle$ is the zero temperature occupation of a quasiparticle with quantum number $\gamma$.
We can now write
\begin{eqnarray}
\langle N|&&
\sum_{M_- } 
\frac{{\cal V}^+ |M_-\rangle\langle M_- |{\cal V}^- }{E_{M-}-E_0}|N'\rangle =
\sum_{m_-}\sum_{\alpha\alpha',{\bf i}{\bf i'}}
\left[1-n_f(\alpha)\right]
\crcr
&& \times
\frac{V_{\alpha,{\bf i}}^*V_{\alpha',{\bf i'}}}{E_{m_-}-E_0 +\epsilon_{\alpha'}}
\langle \psi_f(0)| f_{\alpha}f^{\dagger}_{\alpha'}|\psi_{f'}(0)\rangle
\crcr
&&\quad \times
\langle n |d^{\dagger}_{{\bf i}}|m_-\rangle
\langle m_- |d_{{\bf i'}}|n'\rangle 
\label{omem}.
\end{eqnarray}
A similar expression can be obtained for the matrix elements involving states $|M_+\rangle$,
\beqa
\sum_{M_+ }&& 
\langle N|\frac{{\cal V}^- |M_+\rangle\langle M_+ |{\cal V}^+ }{E_{M+}-E_0}|N'\rangle=
\sum_{m_+}\sum_{\alpha\alpha',{\bf i}{\bf i'}}
n_f(\alpha) 
\crcr
&&\times
\frac{V_{\alpha,{\bf i}}V^*_{\alpha',{\bf i'}}}{E_{m_+}-E_0-\epsilon_{\alpha}} 
\langle \psi_f(0)| f^{\dagger}_{\alpha}f_{\alpha'}|\psi_{f'}(0)\rangle
\crcr
&& \qquad \times
\langle n |d_{{\bf i}}|m_+\rangle
\langle m_+ |d^{\dagger}_{{\bf i'}}|n'\rangle .
\label{omep}
\eeqa
Now, it is straightforward to show that the addition of Eqs. (\ref{omem})-(\ref{omep}) leads to the final expression 
(\ref{hteffect0}).

\section{Tunneling transition rates\label{appendixB}}
As stated in the Sec.~\ref{rates}, the cotunneling transition rates can be calculated applying the Fermi Golden Rule to the effective tunneling Hamiltonian ${\cal H}_{cotun}$. Introducing the density of states $\rho_{\eta\sigma}$ and using Eqs. (\ref{omem}-\ref{omep}), the transition rate from a state $n$ of the central island to an state $n'$, with the transport electron going from electrode $\eta$ to $\eta'$ and its spin  from $\sigma$ to $\sigma'$, are  given by
\begin{widetext}
\beqa
W_{n,n'}^{\eta\sigma\eta'\sigma'}&=&\frac{2\pi }{\hbar} \int d\epsilon \;
\rho_{\eta\sigma}(\epsilon)\rho_{\eta\sigma'}(\epsilon+\Delta_{nn'}) 
f(\epsilon-\mu_\eta)\left(1-f(\epsilon+\Delta_{nn'}-\mu_{\eta'})\right)
\crcr
&& \times
\left|\langle n|\hat{{\cal O}}^{(+)}_{\eta\sigma,\eta'\sigma'}(\epsilon,\epsilon+\Delta)|n'\rangle
-\langle n'|\hat{{\cal O}}^{(-)}_{\eta'\sigma',\eta\sigma}(\epsilon+\Delta,\epsilon)|n\rangle\right|^2,
\label{gamma}
\eeqa
\end{widetext}
with $\Delta_{nn'}=E_n-E_{n'}$ and $f(\epsilon)$ the Fermi-Dirac distribution. The matrix elements of the $\hat{{\cal O}}^{(\pm)}$ operators in Eq. (\ref{gamma}) are defined as
\beqa
\langle n|\hat{{\cal O}}^{(+)}_{\eta\sigma,\eta'\sigma'}(\epsilon,\epsilon')|n'\rangle=\sum_{ii',m_+}
\frac{V_{\eta,i}(\epsilon) V_{\eta',i'}^*(\epsilon')}{E_{m_+}-E_0-\epsilon}\gamma_{nn'}^{m_+}(ii',\sigma\sigma')
\nonumber
\eeqa
and
\beqa
\langle n|\hat{{\cal O}}^{(-)}_{\eta\sigma,\eta'\sigma'}(\epsilon,\epsilon')|n'\rangle=\sum_{ii',m_-}
\frac{V_{\eta,i}^*(\epsilon)
V_{\eta',i'}(\epsilon')}{E_{m_-}-E_0+\epsilon'}\gamma_{nn'}^{m_-}(ii',\sigma\sigma'),
\nonumber
\eeqa
where we have used a simplified notation $V_{\eta,i}(\epsilon)\equiv V_{k(\epsilon)\eta,i}$.These transitions rates are in perfect agreement with the rates obtained by a second order truncation of the $T$-matrix.\cite{Hansen_Mujica_nanolett_2008,Roch_Vincent_arXiv_2011}


\end{document}